\documentclass{PoS}

\usepackage{lineno}

\usepackage[gen]{eurosym}  
\usepackage{changepage} 

\graphicspath{{figs/}}

\title{FlashCam: a fully-digital camera for the medium-sized telescopes of the Cherenkov Telescope Array}

\ShortTitle{FlashCam: a fully-digital camera for CTA}

\author{\speaker{G.~P{\"u}hlhofer}$^{b}$, C.~Bauer$^{a}$, S.~Bernhard$^{g}$, M.~Capasso$^{b}$, S.~Diebold$^{b}$, F.~Eisenkolb$^{b}$, D.~Florin$^{c}$, C.~F{\"o}hr$^{a}$, S.~Funk$^{d}$, A.~Gadola$^{c}$, F.~Garrecht$^{a}$, G.~Hermann$^{a}$, I.~Jung$^{d}$, O.~Kalekin$^{d}$, C.~Kalkuhl$^{b}$, J.~Kasperek$^{e}$, T.~Kihm$^{a}$, R.~Lahmann$^{d}$, A.~Manalaysay$^{c}$, A.~Marszalek$^{f}$, M.~Pfeifer$^{d}$, P.J.~Rajda$^{e}$, O.~Reimer$^{g}$, A.~Santangelo$^{b}$, T.~Schanz$^{b}$, T.~Schwab$^{a}$, S.~Steiner$^{c}$, U.~Straumann$^{c}$, C.~Tenzer$^{b}$, A.~Vollhardt$^{c}$, Q.~Weitzel$^{a}$, F.~Werner$^{a}$, D.~Wolf$^{c}$, K.~Zietara$^{f}$, 
for the CTA Consortium\footnote{Full consortium author list at http://cta-observatory.org}\\
        E-mail: \email{Gerd.Puehlhofer@astro.uni-tuebingen.de}

{\footnotesize
$^{a}$ Max-Planck-Institut f{\"u}r Kernphysik, P.O. Box 103980, D~69029 Heidelberg, Germany;\\
$^{b}$ Institut f{\"u}r Astronomie und Astrophysik, Abteilung Hochenergieastrophysik, Kepler Center for Astro and Particle Physics, Eberhard Karls Universit{\"a}t, Sand 1, D~72076 T{\"u}bingen, Germany;\\
$^{c}$ Physik-Institut, Universit{\"a}t Z{\"u}rich, Winterthurerstrasse 190, 8057 Z{\"u}rich, Switzerland;\\
$^{d}$ Physikalisches Institut, Friedrich-Alexander Universit{\"a}t Erlangen-N{\"u}rnberg, Erwin-Rommel-Str. 1, D~91058 Erlangen, Germany;\\
$^{e}$ AGH University of Science and Technology, Al. Mickiewicza 30, 30-059 Krakow, Poland;\\
$^{f}$ Astronomical Observatory, Jagiellonian University, ul. Orla 171, 30-244 Krakow, Poland;\\
$^{g}$ Institut f{\"u}r Astro- und Teilchenphysik, Leopold Franzens Universit{\"a}t Innsbruck, Technikerstrasse 25/8, A~6020 Innsbruck, Austria}
}

\abstract{The FlashCam group is currently preparing photomultiplier-tube based cameras proposed for the medium-sized telescopes (MST) of the Cherenkov Telescope Array (CTA). The cameras are designed around the FlashCam readout concept which is the first fully-digital readout system for Cherenkov cameras, based on commercial FADCs and FPGAs as key components for the front-end electronics modules and a high performance camera server as back-end. This contribution describes the progress of the full-scale FlashCam camera prototype currently under construction, as well as performance results also obtained with earlier demonstrator setups. Plans towards the production and implementation of FlashCams on site are also briefly presented.}

\FullConference{The 34th International Cosmic Ray Conference,\\
		30 July--6 August, 2015\\
		The Hague, The Netherlands}

\begin{document}

\section{Introduction}
\label{sec:intro}

The Cherenkov Telescope Array (CTA) is the next-generation ground-based Cherenkov telescope facility currently under preparation by an international consortium \cite{bib:Design2011}. CTA will consist of several tens of telescopes of different sizes, to cover the photon energy range from few tens of GeV to few hundreds of TeV. 
The FlashCam group has developed a camera concept for CTA telescopes, that is based on purely digital processing of the photosensor signals and Ethernet-based front-end readout. Over the last years, the concept has been verified against CTA requirements in several evaluation setups. Currently (2015), the group is constructing a full-scale camera prototype for CTA medium-sized telescopes (MSTs). In parallel, preparations are ongoing towards production and implementation of FlashCam MST cameras for the final array.

\begin{figure}[t]
    \centering
	\includegraphics[width=1.0\textwidth]{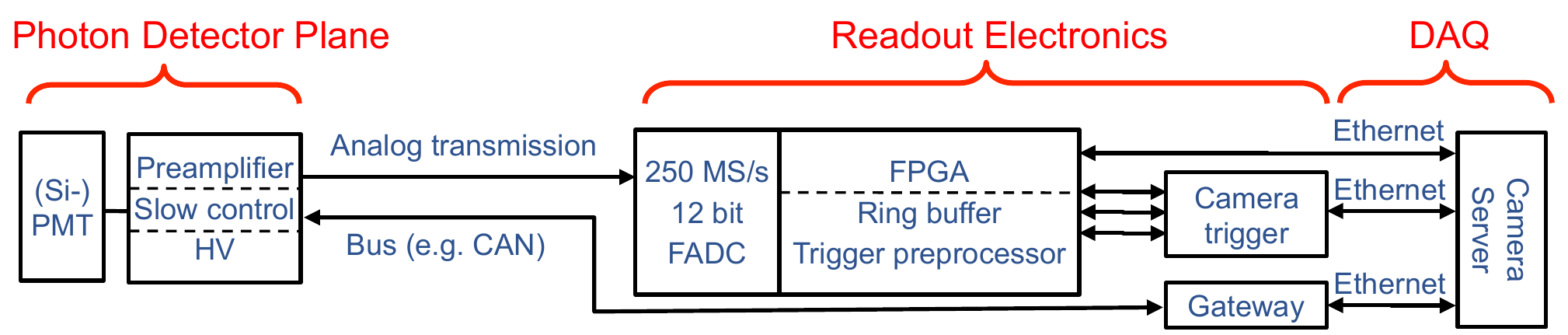}
	\caption{Basic building blocks of the FlashCam signal chain.}
	\label{FCFigArchitectureFC}
\end{figure}

\section{The FlashCam Concept and the Design for an MST camera}
\label{sec:concept}

FlashCam is a self-contained, ready to use camera for Cherenkov telescopes. It includes the following sub-systems: \vspace{0.15cm}

\setlength{\leftskip}{20pt}
\setlength{\parindent}{-9pt}

	$\bullet$ Photon detection system using high quantum efficiency PMTs;

	$\bullet$ Trigger and digitization electronics based on a fully digital approach;

	$\bullet$ Ethernet-based camera data acquisition (DAQ) system, based on ``off the shelf'' components, including DAQ computer for camera event building (``camera server'');

	$\bullet$ Camera mechanics and cooling system;

	$\bullet$ FlashCam calibration and monitoring system;

	$\bullet$ Auxiliary systems for slow control, monitoring, and safety;

	$\bullet$ High performance software suited for high bandwidth (GByte/sec) front-end readout based on Ethernet standard off the shelf components \cite{bib:RawEthernet}. \vspace{0.15cm}

\setlength{\parindent}{20pt}
\setlength{\leftskip}{0pt}

The design of FlashCam follows a ``horizontal'' architecture, with the photon detector plane (PDP), the readout electronics (ROS), and the data acquisition system (DAQ) as key building blocks (see Fig.~\ref{FCFigArchitectureFC}). The PDP contains photomultiplier tubes arranged in a hexagonal structure with 50 mm pixel spacing. 12 PMTs are combined in a PDP module, which provides high voltage, and contains pre-amplifiers, as well as a micro controller for slow control, monitoring, and safety functions. The signals are then transmitted via cables to the readout electronics, the design of which is based on a fully digital approach with continuous signal digitization. The signals are digitized continuously with $12$-bit FADCs with a sampling frequency of 250 MS/s. The digitized signals are processed and buffered on FPGAs, which at the same time are used to derive the camera trigger based on the digitized signals. All electronics modules are read out directly via a camera-internal, high performance Ethernet-network, using off-the shelf switches and a standard commercial computer. On this ``camera server'', custom developed software is implemented for a high performance front-end to back-end data transfer and to do e.g.\ the event building, optional zero-suppression, event selection, extraction of array trigger information, and data formatting. It provides interfaces to the CTA-wide array control, data acquisition, and to the software-based array trigger. 

\begin{figure}[t]
    \centering
	\includegraphics[trim=0cm 0cm 0cm 1.75cm, clip=true, width=0.7\textwidth]{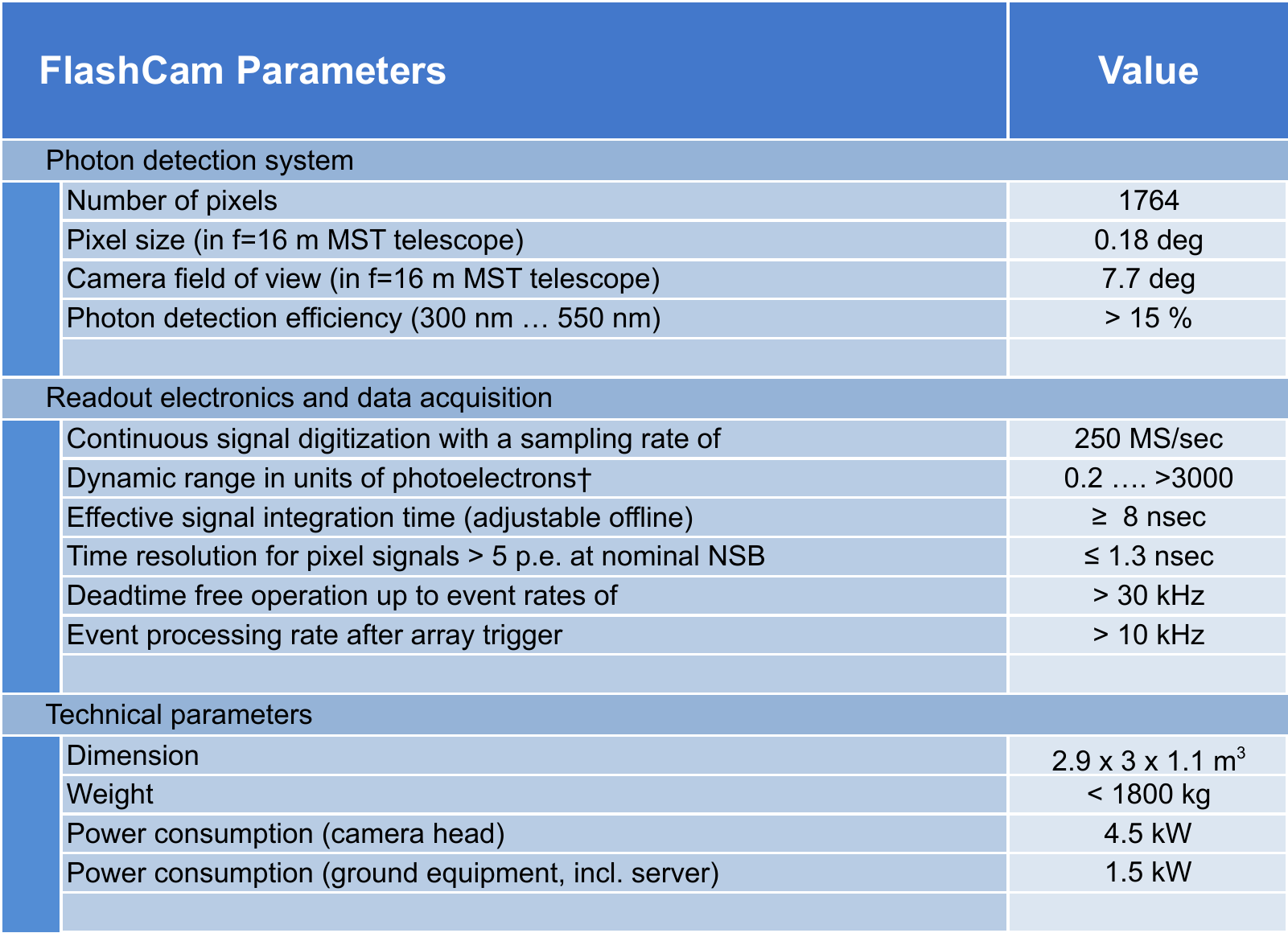}
	\caption{FlashCam for MST key performance parameters. $\dagger$: with an amplitude resolution better than CTA requirements.}
	\label{FlashCamSpecsSummary}
\end{figure}

Fig.~\ref{FlashCamSpecsSummary} lists key performance parameters of the camera. Amongst the highlight features of the readout system are:

\setlength{\leftskip}{20pt}
\setlength{\parindent}{-9pt}
    $\bullet$ Firmware-controlled functionalities; multi-boot option for different firmware versions and external Ethernet upload of new firmware;

	$\bullet$ High, sustained input data rate capabilities at the camera server of $>2$\,GByte/s, including event building and data formatting;

	$\bullet$ Dead-time free operation up to burst event rates of $>50$\,kHz (e.g. due to short-lived artificial light sources) and sustained rates of $>30$\,kHz, with transmission of digitized signal traces to the camera server;

	$\bullet$ Flexible trigger scheme: different algorithms can be configured and parametrized depending on observation conditions;

	$\bullet$ Delayed external trigger capability, allowing to read out telescopes in array-triggered events, even if the camera did not trigger by itself, or upon a second, lower-threshold trigger, due to the $32\,\mu$s dead-time free ring buffer for FADC data.

\setlength{\parindent}{20pt}
\setlength{\leftskip}{0pt}

\begin{figure}[t]
    \centering
    \includegraphics[trim=0cm 0cm 0cm 0.1cm, clip=true, width=0.63\textwidth]{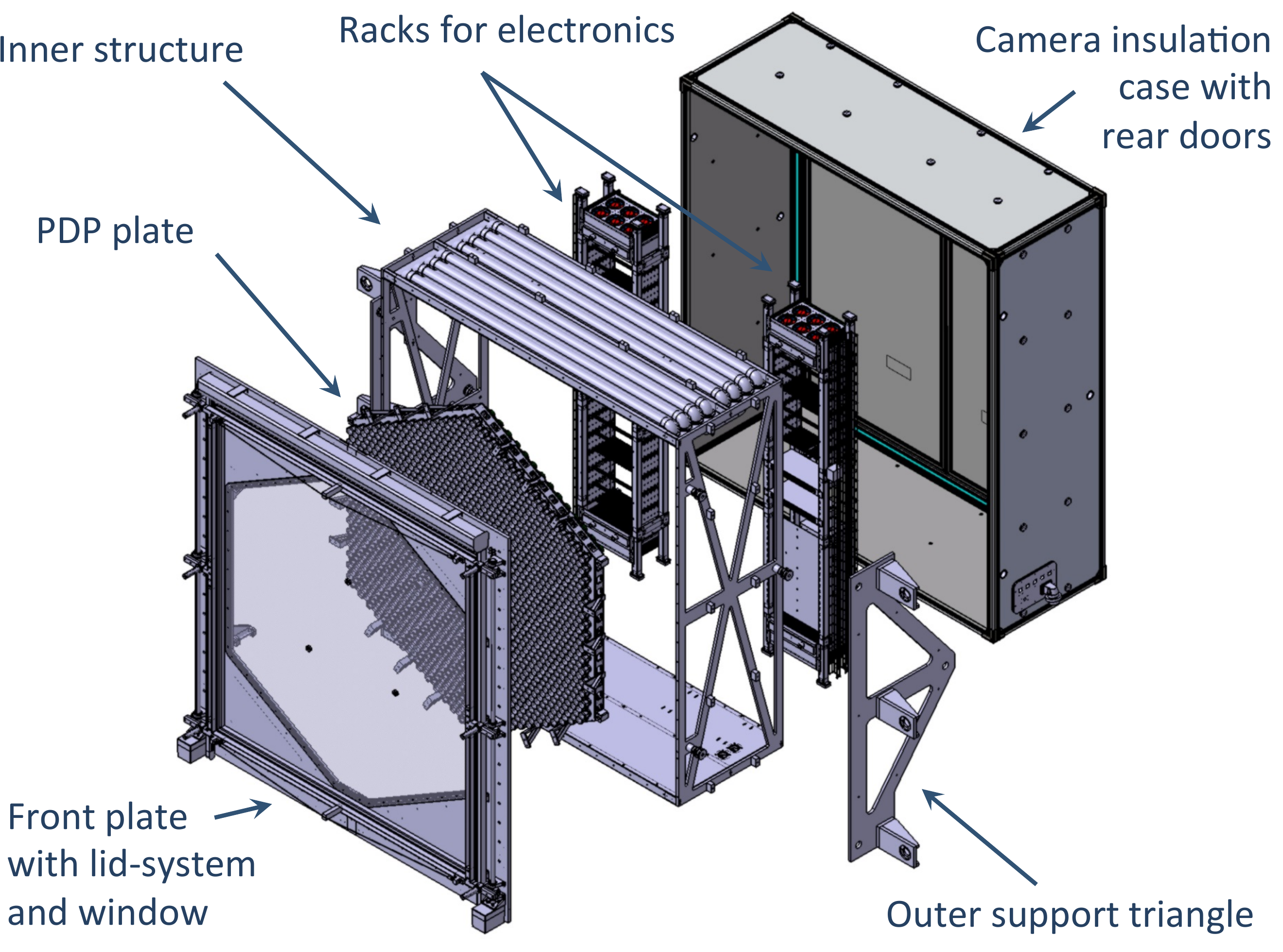}
   \caption{Exploded view of the FlashCam MST camera.}
   \label{FCFigCameraExplosion}
\end{figure}
The PDP and the readout system with Ethernet switches are physically contained in the mechanical camera body in the focal plane region of the MST telescope, while the camera server is planned to be located inside the central control building of the telescope array. Fig.~\ref{FCFigCameraExplosion} shows a schematic view of the FlashCam camera body, with the mechanical structure and thermal insulation, the photon detector plane (PDP plate), and the rack system for the readout electronics.

\section{The full-scale FlashCam Prototype Camera}
\label{The full-scale FlashCam Prototype Camera}

\begin{figure}[t]
 	\centering
 	\includegraphics[width=0.3\textwidth]{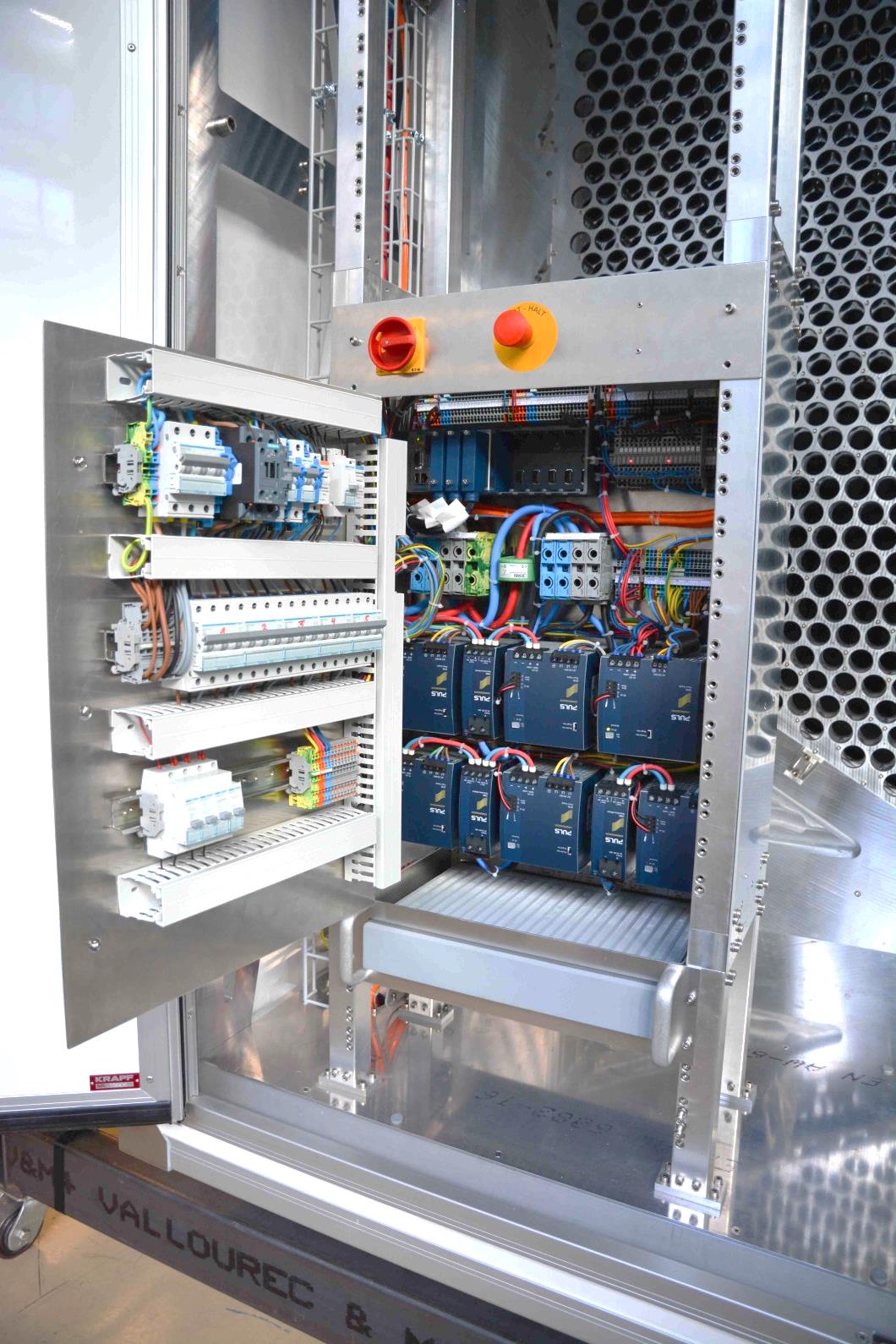}
    \qquad
  \includegraphics[width=0.3\textwidth]{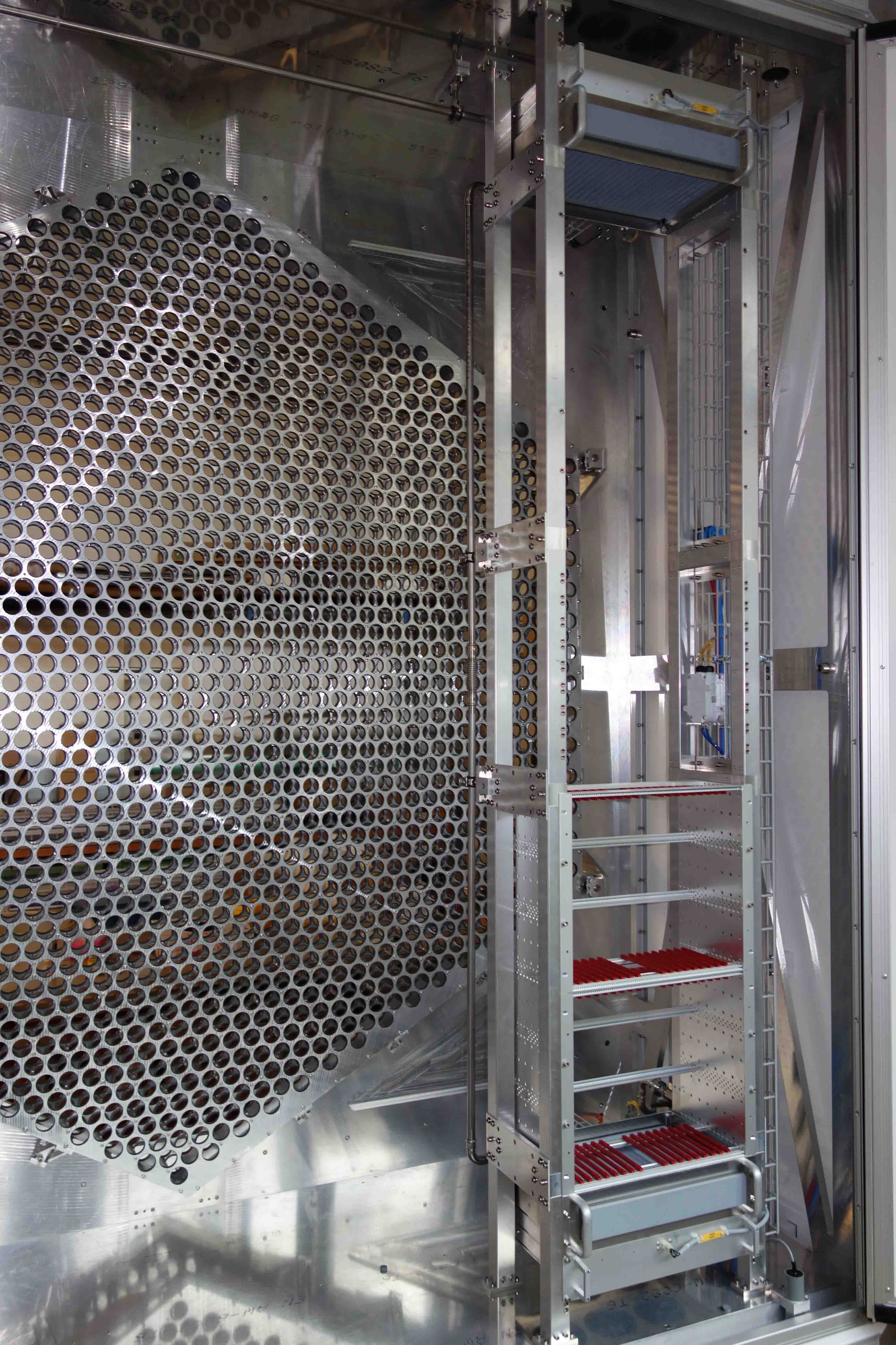}
    \qquad
  \includegraphics[width=0.137\textwidth]{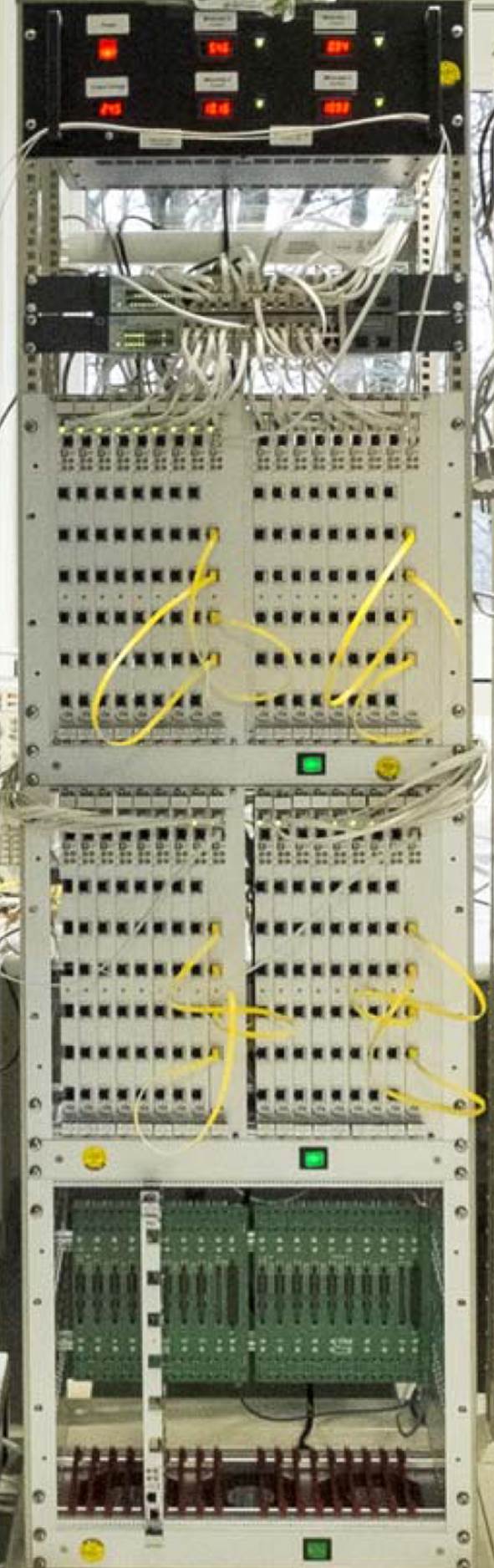}
	\caption{Photos of the FlashCam prototype camera. \textbf{Left}: Safety control cabinet installed inside the camera body. \textbf{Middle}: rear view into the camera body showing one of the installed electronics racks and an attached cooling pipe. \textbf{Right}: Rack with readout electronics under test.}
	\label{FCFigControlCabinetCameraRear}
\end{figure}

Activities of the FlashCam team are currently centered around the finalization of a full-size FlashCam camera prototype in 2015. 
The production of the camera mechanics and electronics is well advanced. 
Fig.\,\ref{FCFigControlCabinetCameraRear} gives an impression of some prototype subcomponents. In the following, the status of the camera prototype as of mid 2015 is sketched.

\subsection{The camera body and safety control} \label{The camera body and safety control}

The camera structure is based on a load-carrying frame that is connected to the telescope mast structure via its side plates. The camera body surfaces are made from lightweight aluminium-foam composite which also serves as thermal insulation. The front lid has passed a first set of endurance tests, the completed body has undergone environmental tests such as simulated hail impact.

Inside the camera, the 19'' racks housing the readout electronics have been installed on the sides, permitting installation and later servicing of the electronic boards. The fluid (water/Glycol) cooling system has been installed and underwent first testing, final commissioning and thorough testing of the cooling and ventilation system is currently ongoing. A preliminary plexiglas entrance window will serve to permit verification of the camera's temperature behavior in the final near-sealed configuration. The camera safety controller has been installed and commissioned, corresponding control software on micro controller level has been developed, and an expert mode for direct human interfacing is available for the continued tests.

\subsection{Photon detectors and photon detector plane electronics}\label{Photon detectors and photon detector plane electronics}

Photon detectors and readout electronics for the prototype camera are ordered in two batches of $\gtrsim 800$ channels each. The first batch has been ordered and delivered. 400 channels will be equipped with the baseline Hamamatsu $8$-dynode CTA PMTs (R11920-100), the remaining $400$ channels with a recently developed Hamamatsu $7$-dynode PMT (R12992-100). PDP PCBs have been developed and ordered for these two variants accordingly. All PMTs underwent income control tests and were assembled on the PCBs. As of now, the tests have shown that both PMT types perform as expected. PDP modules are currently undergoing thorough testing before integration into the camera prototype. The slow control software controlling the PDP via CAN bus will be finalized in the upcoming months, using a set of specifically designed PDP emulator boards.

\subsection{Crate-based readout electronics}\label{Crate-based readout electronics}

The readout electronics is segmented into three sectors of $\sim$ 600 channels each. FADC readout electronics for 840 channels, as well as backplanes and trigger interface cards for a full 1764-channel camera have already been delivered before end of 2014, and have undergone successful acceptance tests. Camera segments are being integrated and tested from smaller scales up to full camera scale (with up to 50\% of the channels fully equipped), to verify analog performance and digital signal communication on all relevant camera-size scales, e.g.\ between several fully-equipped crates and across camera sector boundaries. Orders to have the remaining electronics and PMTs for a full $1764$-pixel camera ready by the second half of 2015 are currently being prepared.

\subsection{Readout firmware and software, analysis framework}\label{Readout firmware and software, analysis framework}

Firmware for the readout of the FlashCam FADC modules has been developed and used over the past years for the different stages of the development and evaluation boards, which were initially based on the Spartan 3 and later on the final Spartan 6 FPGA. For the final cameras, a redesigned production version of this firmware is nearly finished. A common analysis library is being developed and tested with data from the FlashCam 144-pixel ``mini-camera''.

\section{Camera Calibration and Performance Verification}

Scientific performance aspects of the FlashCam camera have been verified with demonstrator hardware including a FlashCam 144-pixel ``mini-camera'' over the last years already. To test and verify the system under realistic conditions, the preamplifier/FADC properties have been characterised using an artificial light source (PicoQuant LDH-P-C405 laser head) to simulate light received from air showers and a DC light bulb to simulate the night sky background (NSB). The amplitude of the light source was modified using two filter wheels to avoid any change of the laser characteristics when changing the laser output. The filter wheels have been calibrated to an accuracy of around $1$\,dB ($10$\%). The measurements were performed with the PDP module equipped with CTA PMTs from Hamamatsu (R11920). 

Key parameters that were verified are the dynamic range and amplitude resolution of measurements on pixel level, as well as the timing resolution. The required dynamic range is $1$\,p.e.\ to $1000$\,p.e., with an additional goal to extend the range to $2000$\,p.e., at NSB rates of $125$\,MHz. While the design of FlashCam is optimised for this dynamic range, it is in fact capable of processing up to $5000$\,p.e.\ according to measurements. Charge resolution results show compliance both with the CTA resolution requirements and goals. An additional CTA camera design goal is to achieve a pixel arrival time resolution of better than $2$\,ns above $5$~p.e.\ under normal NSB conditions. The measured time resolution is about $1.3$\,ns for a signal amplitude of $5$\,p.e.\ and a NSB of $240$\,MHz, fulfilling the goal. On system level, the performance and stability of the digital trigger communication (with a total transmission capacity per camera of 2.7\,Tbit/s) has been verified on individual crate level. Results of these performance verifications have been reported e.g. in \cite{bib:Gamma2012} and \cite{bib:SPIE2014}.

\begin{figure}[tp]
    \centering
    \includegraphics[width=\textwidth]{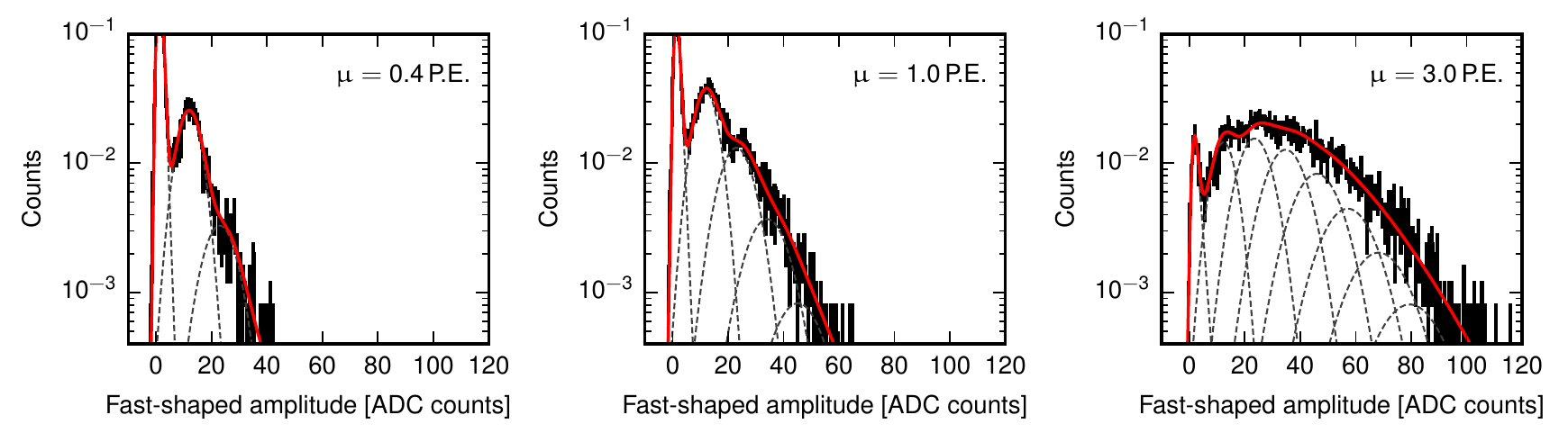}
   \caption{Amplitude spectra for three levels of illumination ($0.4$~p.e., $1.0$~p.e., and $3.0$~p.e.\ on average) for one pixel of the $144$-pixel prototype. Maximum likelihood fits are shown as red lines, the individual peaks as black dashed lines.}
   \label{FCFigPESpectra}
\end{figure}

Tests of the full-size prototype on system level with close-to-final configuration will serve to verify full system behavior and performance. A preliminary version of the calibration device that will be installed at each telescope is under development and will be included in the test suite. This calibration device will consist of a pulsed light source similar to the laboratory setup described above. The device will serve for initial calibration and tracking of the calibration constants (i.e., pedestals and factors for converting ADC counts to photoelectron (p.e.) counts). The extraction of calibration parameters from single photoelectron spectra has been tested with measurements from the $144$-pixel camera prototype. In Fig.~\ref{FCFigPESpectra}, maximum likelihood fits to photoelectron spectra obtained at different light intensities are shown for one pixel. The calibration parameters obtained at varying illumination levels agree to better than $10$\% for average illumination levels up to $\sim 5$~p.e. Data taken with a calibrated light attenuator is used to match the analysis of the saturation region to the linear region. The uncertainty of this method is expected to be in the range of $5$--$10$\%.

Another scope of the camera prototype is to further evaluate reliability and lifetime of the individual components, to verify maintenance plans and spare policy, and possibly to improve the design if necessary. Experience with the existing electronics has proven already high reliability of the components. In total, $> 250$ FADC channels, $> 30$ FADC boards and a $144$-pixel PDP  have been operated since 2013 for in total more than (estimated) $1000$ hours, including numerous power cycles. First stress tests and performance measurements with a crate equipped with 196 readout channels in a climate chamber have already been performed with several tens of cycles in the $10^{\circ}\mathrm{C}$ to $40^{\circ}\mathrm{C}$ temperature regime, as well as at temperatures of $> 70^{\circ}\mathrm{C}$ for several hours. No performance loss or loss of modules has been observed yet.

Tests such as highly accelerated lifetime tests and stress tests (temperature and humidity cycles) will continue with the fully-equipped camera prototype e.g. to obtain meaningful failure rates, and also to continue performing failure mode analysis and to develop corresponding procedures.

\section{Plans towards production}

\begin{figure}[t]
    \centering
    \includegraphics[trim=0cm 0.2cm 0cm 0.2cm, clip=true, width=1\textwidth]{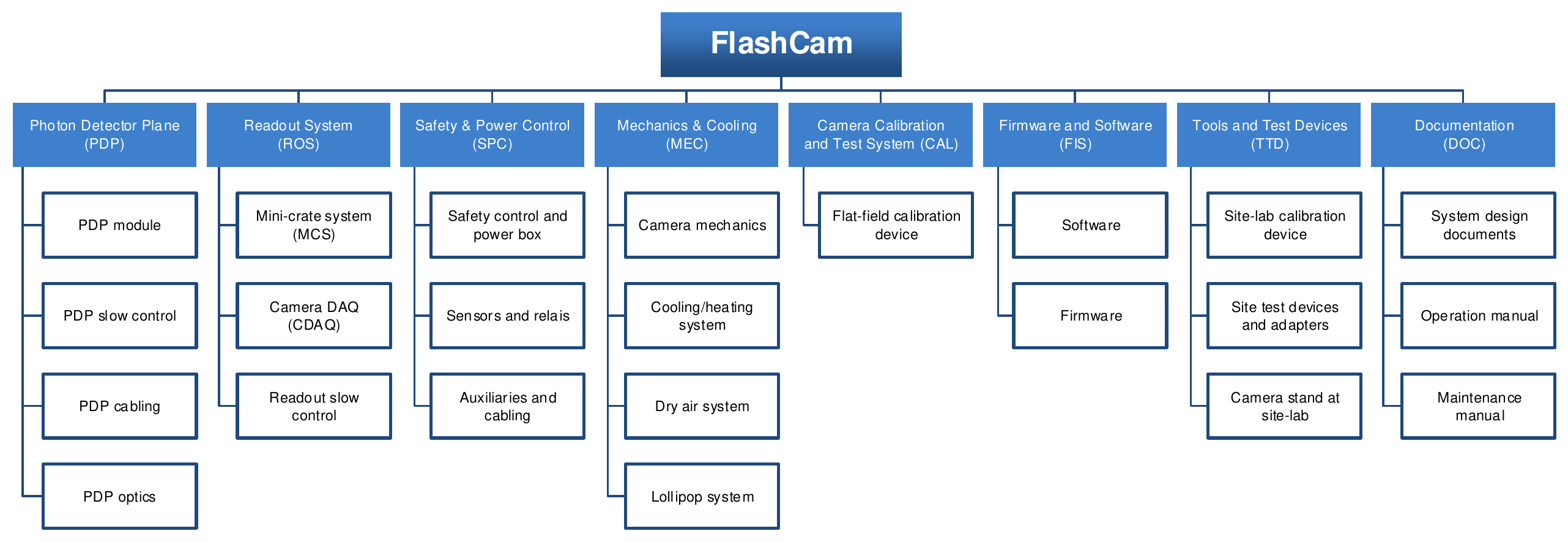}
   \caption{FlashCam product breakdown structure down to level 2.}
   \label{FCFigPBSDiagram}
\end{figure}

The FlashCam team is preparing for the production of cameras for the MST telescopes of CTA. As described above, a full-size prototype camera to verify the design is currently being built. During 2016 pre-production of two cameras should start to optimize production processes and to prepare for mass production. Planning of the production process is currently ongoing, some aspects are briefly sketched below.

\subsection{Manufacture and assembly} \label{Manufacture and assembly}

FlashCam camera production is envisaged to take place in a similar style as it is done in large high energy physics experiments. Production of components and sub-assemblies, especially of large quantity items, will take place in industry. Assembly and test of camera sub-systems (like readout electronics crates, photon detector plane modules, mechanics, slow-control systems, etc.) will be done at the contributing laboratories. The pre-fabricated and tested sub-systems of the camera will then be shipped to one or two central laboratory(s), where camera integration, system tests, and calibration will take place, before shipment to site. In-house production of components will be envisaged only if there are special requirements that are not easily fulfilled in industrial production processes, or if it will be significantly more cost efficient when done in-house. So far, no significant in-house production of components is foreseen.

\subsection{Maintenance} \label{Maintenance}

Besides the delivery and commissioning of fully functional cameras, detailed maintenance plans are being worked on to enable the observatory to run FlashCams successfully in the forthcoming years. Most of the components of FlashCam are maintenance-free. One of the design goals of the FlashCam project has been to reduce and simplify the remaining necessary maintenance effort as well as the required repair and exchange work, in comparison to current-generation cameras. The implementation therefore follows to a large extent the concept of line-replaceable units (LRU). It is targeted that it should be possible to replace any active component such as PDP modules or electronics boards of the ROS in less than one hour, i.e.\ from the moment where a person arrives at the secured telescope until the camera is ready for operation again. Therefore the camera body is designed as a ``walk-in'' camera. All PDP electronics modules are accessible from inside of the camera and in case of a replacement of modules there is no need to touch or remove the delicate optical system of the camera or the camera window. All readout electronics boards can even be accessed by just opening the rear doors of the camera and without entering the camera body.

Most importantly, all routine maintenance work or exchange of modules can be done with the camera being installed at the telescope, and it is not necessary to take the camera to a workshop for this type of work.

\section{Summary}
\label{sec:summary}  

The FlashCam team has demonstrated that a fully-digital camera readout concept is suitable for classical PMT-based Cherenkov telescope cameras. Full verification of the concept against simulations and CTA requirements has been performed with demonstrator setups and a 144-pixel mini-camera setup. A full-scale prototype camera for the CTA MST is currently under construction. Plannings for large-scale production are ongoing.

\subsection*{Acknowledgements}

We gratefully acknowledge support from the agencies and organizations under Funding Agencies at \href{www.cta-observatory.org}{www.cta-observatory.org}.

\end{document}